# Event Vision Sensor: A Review

Xinyue Qin, Junlin Zhang, Wenzhong Bao, Chun Lin and Honglei Chen

***Abstract*—The emergence of event-based vision sensors is rooted in neuromorphic engineering, with an initial design concept aimed at mimicking the functionality of rod cells in the eye to achieve high dynamic range and temporal differential sensing. By monitoring temporal contrast, event-based vision sensors can provide high temporal resolution and low latency while maintaining low power consumption and simplicity in circuit structure. These characteristics have garnered significant attention in both academia and industry. In recent years, the application of back-illuminated (BSI) technology, wafer stacking techniques, and industrial interfaces has brought new opportunities for enhancing the performance of event-based vision sensors. This is evident in the substantial advancements made in reducing noise, improving resolution, and increasing readout rates. Additionally, the integration of these technologies has enhanced the compatibility of event-based vision sensors with current and edge vision systems, providing greater possibilities for their practical applications. This paper will review the progression from neuromorphic engineering to state-of-the-art event-based vision sensor technologies, including their development trends, operating principles, and key features. Moreover, we will delve into the sensitivity of event-based vision sensors and the opportunities and challenges they face in the realm of infrared imaging, providing references for future research and applications.**

## I. Introduction

"Like fat free milk, event-based silicon retinas can free the consumer from consumption of excess energy.", that is how Tobi Delbruck introduced the event vision sensor (EVS) feature in "Neuromorphic Vision Sensing and Processing" [1].

In the last decade, high spatial resolution frame cameras have advanced rapidly, achieving pixel pitch sizes smaller than those of the human eye's photoreceptors [2]. These sensors capture vast amounts of data, driving the development of technologies such as convolutional neural networks (CNNs) and graphics processing units (GPUs) to process this information efficiently [3]. However, these focal plane arrays (FPAs) are constrained by scene-independent, preset integration times. Additionally, as pixel sizes decrease, the number of photons received by each pixel diminishes, leading to longer exposure times and reduced frame rates [4]. To address the challenges of limited frame rates and high latency in FPAs, technologies such as region of interest (ROI) imaging have been developed, allowing selective processing of specific areas to improve image performance.

Manuscript received XXX; revised XXX; accept XXX. Date of publication XXX; date of current version XXX. This work was supported by the XXXX. (Corresponding authors: Honglei Chen.)

Xinyue Qin, Junlin Zhang, Chun Lin and Honglei Chen are with the Key Laboratory of Infrared Imaging Materials and Detectors, Shanghai Institute of Technical Physics, Chinese Academy of Sciences, Shanghai 200083, China (qinxinyue22@mails.ucas.ac.cn;zhangjl@mail.sitp.ac.cn;chun_lin@mail.sitp.ac.cn; henhl@mail.sitp.ac.cn)

Wenzhong Bao is with the School of Microelectronics, Fudan University, Shanghai 200433, China (baowz@fudan.edu.cn)

EVS responds asynchronously to scene changes and generates sparse, precisely time-encoded differential data, at a low price of computational complexity and power consumption. For certain tasks, such as object tracking or recognition, only sparse edge information is needed, whereas the large amount of data captured by frame cameras is often redundant [5]. By extracting key edge features, not only can redundant data be significantly reduced, but transmission and processing speed can also be greatly increased. Although some methods like frame differential methods [6] and pixel ADCs [7] have been developed, they still rely on similar paradigms of conventional frame cameras, still trapping in fixed frame rates and preset integration times. These limitations make them less effective for capturing high-speed or dynamic scene changes and lead to unnecessary data redundancy, ultimately reducing processing efficiency.

The development of EVS initially focused on addressing challenges such as readout rate [8], encoding precision [9] and quantum efficiency (QE) [10]. These early challenges have been partially overcome through innovations in backside illumination technology [11] and readout circuits [12]. In recent years, hybrid [13] and high-resolution [14] EVS have been developed to seamlessly integrate with existing vision systems and chase for highly precise spatial imaging. Various EVS has been designed and continued to adapt to diverse application scenarios, with discussions centered on latency, timestamp accuracy [14], and the readout methods [15] required for high-resolution designs. The field remains on a trajectory of progress, striving for further improvements in contrast sensitivity [16], multi-stream readout capabilities [17], faster readout rates [11], and on-chip event processing [18] to achieve even greater performance.

Key players in this field include tech giants like Omnivision [14], [19], Sony [20] and Samsung [11], alongside start-ups such as Prophesee [21]. Prominent research institutions include the Institute of Neuroinformatics at UZH-ETH Zurich [16] and the Instituto de Microelectrónica de Sevilla (CSIC-USE) [22]. Major undergoing projects in this domain include Visualise [23] and NimbleAI [24] in Europe, as well as the Fast Event-Based Neuromorphic Camera and Electronics (FENCE) [25] initiative in the United States.

Recent reviews in this field have predominantly concentrated on event-based vision [26], processing [27], and various practical applications [10], [11]. However, a comprehensive review summarizing the advanced EVS is still lacking. This review will elaborate on the working principles and characteristics of EVS, provide an overview of the development of event cameras over last decades, and discuss the unique challenges and opportunities associated with EVS in infrared imaging.

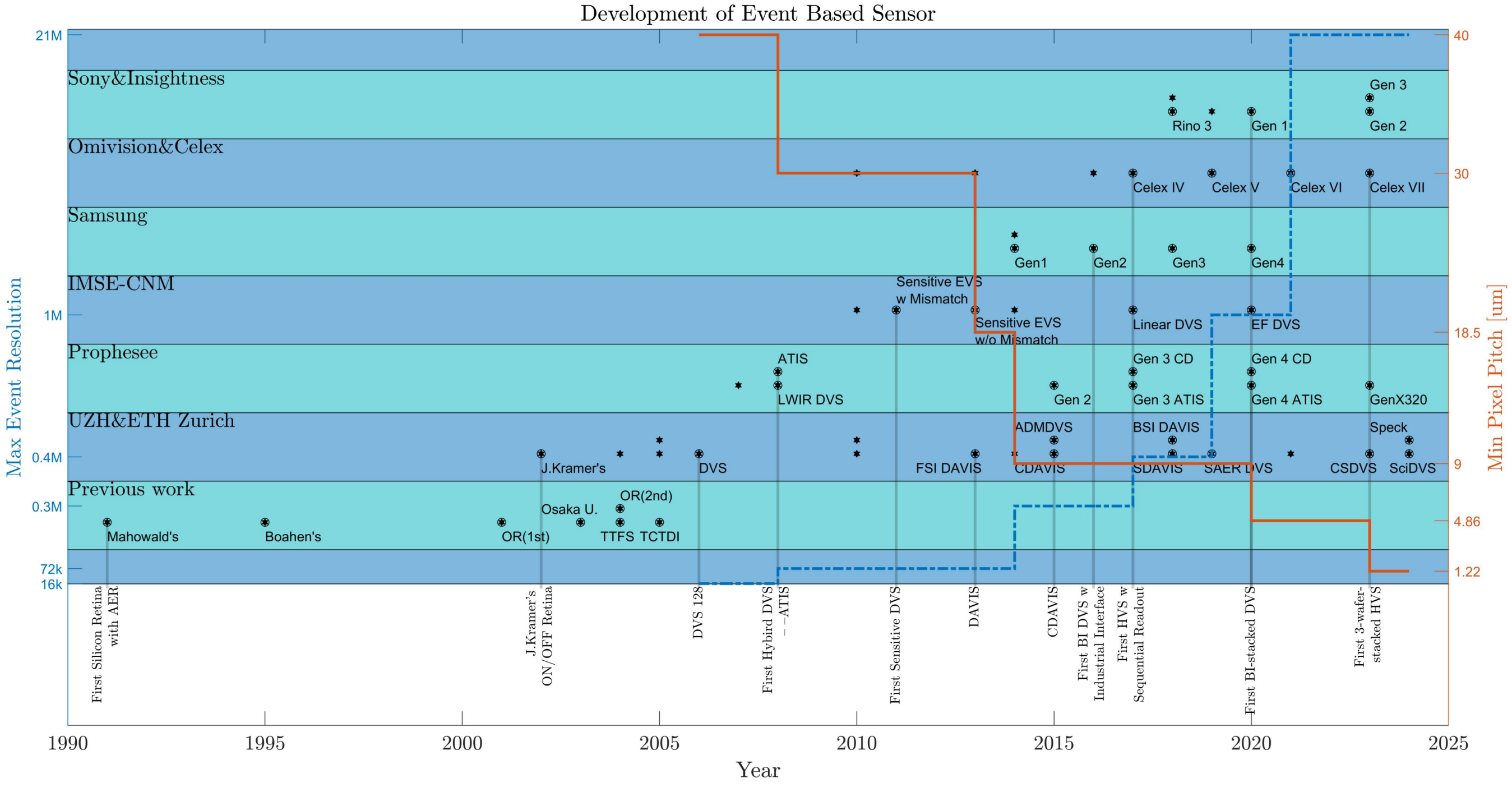

**Fig. 1.** The development of event-based sensor

## II. From Neuromorphic Engineering to Event Camera

Retinal structures in nature exhibit remarkable diversity across species, each adapted to the specific needs of the organism. Biologists have reported on the different retinas of over 50 species [30], [31]. Some animals, such as octopuses and frogs, can only perceive moving objects. Mammals, on the other hand, can detect intensity information across three distinct spectral ranges—red, green, and blue—enabling them to perceive a broad spectrum of light and to process visual information in greater detail [2]. The mantis shrimp has an even more sophisticated visual system, capable of perceiving intensity information across 12 spectral ranges and has evolved structures that can detect polarization [17]. All these creatures have two types of photoreceptor cells in the retina: rod cells and cone cells. Rod cells are sensitive to low-light conditions, providing the ability to detect brightness and facilitating vision in dim environments. Cone cells, on the other hand, are responsible for color vision and are more active in well-lit conditions, allowing these creatures to perceive fine details and discern colors [32].

Neuromorphic engineering, founded by C.A. Mead, aims to explore biological solutions to address the data redundancy and low efficiency inherent in digital processing systems [33]. Many pioneering innovations have emerged from C.A. Mead's group, including silicon retinas, cochleas, neurons, and non-volatile memory [34]. To tackle the off-chip communication challenges of these neuromorphic devices, the Address-Event Representation (AER) communication circuit was initiated by Massimo Sivilotti [35] and further developed by Boahen [50][51]. The first silicon retina with AER, based on the three-layer Kufler retina model, was built by Mahowald during her doctoral research [38].

The EVS photoreceptor stage is designed to mimic rod cells, allowing the retina to function effectively under low-light conditions [2]. The lower the illumination on the rod cells, the more active they become. This feature is imitated by the transduction transistors operating in the weak inversion regime, which allows them to respond to light in a similar way to natural rod cells. Nowadays, comparing to digital signal process system, the robust of analog signal process system is still debatable [39]. However, large number of analogue circuits are still used before the analog signal are encoded to digital signal.

Early neuromorphic vision sensors and active pixel sensor (APS) emerged almost at the same time. Compared to APS these neuromorphic sensors often suffered from large pixel sizes, severe fixed pattern noise, and low fill factors. However, with the evolution of technology, these shortcomings have gradually been overcome, and EVS has demonstrated unique advantages in image processing and computational efficiency.

EVS is well-known as a temporal contrast differential or delta modulation mechanism nowadays, however, there were some pseudo EVS detecting absolute intensity shining on the pixel in the time domain. The intensity was also encoded and communicated in biologically inspired fashions, yet these pixels didn't react to the scene dynamically. One is the so-called "Octopus Retina" (OR) [40] encoding intensity in a rate coding fashion [41] [30], i.e., instantaneous frequency (or inter-spike intervals). The other is the so-called "time-to-first-spike" (TTFS) imager encoding intensity in the latency coding fashion, and intensity information is encoded into one-spike occur time [41] [30].

The first EVS considered with commercial prospects is DVS 128 born in the Convolution AER Vision Architecture for Real-time (CAVIAR) project, whose predecessor is Jorg Kramer's transient sensor computing temporal derivative on rectified charge feedback and ON/OFF comparator circuit imitating to bio-visual process. In contrast to previous bioinspired sensors, serval contributions had been made in DVS128 including:

1) A novel pixel with a real-time logarithmic photosensor and a well-matched self-timed switched-capacitor amplifier [42]
2) Programmable on-chip bias generation circuit [43]
3) A compact camera system with a high-speed USB interface [44]

The nature of the temporal derivative makes DVS no perception of the intensity of light. Hybrid EVS has been developed, whose pixels are designed to integrate the intensity measurement with EVS pixels. The two most representative are asynchronous time-based image sensors (ATIS) and dynamic and active pixel vision sensors (DAVIS).

For the ATIS [10], one of the advantages is the achievement of pixel-level redundancy suppression for intensity information. The real-time intensity information is compressed into logarithm form and measurement can be triggered asynchronously by corresponding events. Additionally, improved SNR and high dynamic range intensity measurement are implemented by introducing time-domain coding pulse width modulation (PWM), with time-domain correlated double sampling. The drawback of ATIS's implementation to measure intensity is as well as introducing another photodiode, which occupies approximately 10% of the pixel area.

DAVIS [12], born in the SeeBetter project, combines DVS with an active pixel sensor. The intensity measurement of DAVIS is a no-compression redundant frame-based output. Without additional photodiodes, DAVIS add a few transistors into the pixel but only increases the 5% of the pixel area. The APS part integrates the photocurrent adapting from the photoreceptor's transistor load on a sampling cap. Except for DAVIS, there are two kinds of prototypes created in this project [45], color DAVIS (CDAVIS) [13] and sensitive DAVIS (SDAVIS) [46]. The CDAVIS first uses pinned photodiodes to achieve low readout and KTC noise [14].

To enhance contrast sensitivity, several efforts on in-pixel design have been made in the early development. The way to realize this is by maximizing overall the voltage gain $A_T = A_v C_1/C_2$, $A_v$: voltage gain of the front-end pre-amplifier [30] [47]. One uses a two-stage capacitive feedback amplifier, but this approach limits EVS frequency response [48]. Photoreceptor variants were designed to improve Av, but these methods either increase sensitivity to transistor mismatch [49] or reduce intra-scene DR [46], [50].

To minimize the loss of information during the encoding stage, the quantization quality of the event encoder is crucial for applications that require precise event accuracy, such as video reconstruction. A new self-timed reset mechanism, asynchronous delta modulation (ADM), has been implemented within the EVS pixel core without the need for a large-area implementation of Digital-to-Analog Converters (DACs) [9]. Unlike traditional feedback-and-reset mechanisms, this feedback-and-subtract approach enhances signal integrity [51].

Early EVS prototypes performed well in asynchronous mode at low resolutions. As EVS resolutions increase, the growing number of pixels detecting changes and generating events simultaneously makes it challenging to assign accurate timestamps [11]. One practical example is when the DVS is in motion, where the event rate can become extremely high. Motion artifacts occur due to the uncertainty and delays introduced by the arbiter [52]. Since the efficiency of the AER readout mode assumes data sparsity, the event output capacity of EVS can become saturated in dense scenes [53]. To address this, the sequence scan readout mode was first introduced in the Celex IV by Omnivision [15] (following its acquisition of Celex). Similar to traditional cameras, this readout method also suffers from decreasing frame rates as resolution increases and output interface bandwidth becomes constrained [20].

To mitigate these limitations, methods such as event drop filters and event compression have been introduced to filter or ignore non-event pixels, as EVS pixels respond only to changes in the scene [14], [54]. Despite the challenges of scan readout mode, arbiter mode offers the advantage of fast response times, particularly in low-activity scenes [18]. Therefore, it remains valuable for low-resolution or low-activity applications, such as IoT, industrial automation, and surveillance.

EVS also faced challenges with limited readout bandwidth and low quantum efficiency (QE). In the earlier stages, when EVS resolution was relatively low, data output could rely on custom parallel interfaces. However, as resolution increased, the mismatch between event generation rates and readout speeds led to the misconception that EVS was only suitable for static scenarios. New CMOS image sensor (CIS) processes and industry-standard interfaces helped overcome this challenge.

Samsung's EVS Gen 2 was the first to implement the MIPI interface and backside illumination (BSI) process, achieving a maximum output bandwidth of 300 Mbps [11]. Two DAVIS sensors with identical pixel sizes and circuits were fabricated using BSI and frontside illumination (FSI) processes. Compared to the FSI DAVIS, the BSI DAVIS exhibited significant improvements, including a four-fold increase in sensitivity [55]. Other notable advancements included a boost in fill factor from 22% to nearly 100% and a rise in peak QE from 24% to 93%. QE performance was also greatly improved across the ultraviolet (UV) range, with a 10x increase, and in the near-infrared (NIR) range, with a 3x improvement. However, the BSI process also introduced challenges, such as increased crosstalk and parasitic photocurrent, which led to higher leak noise events [56] .

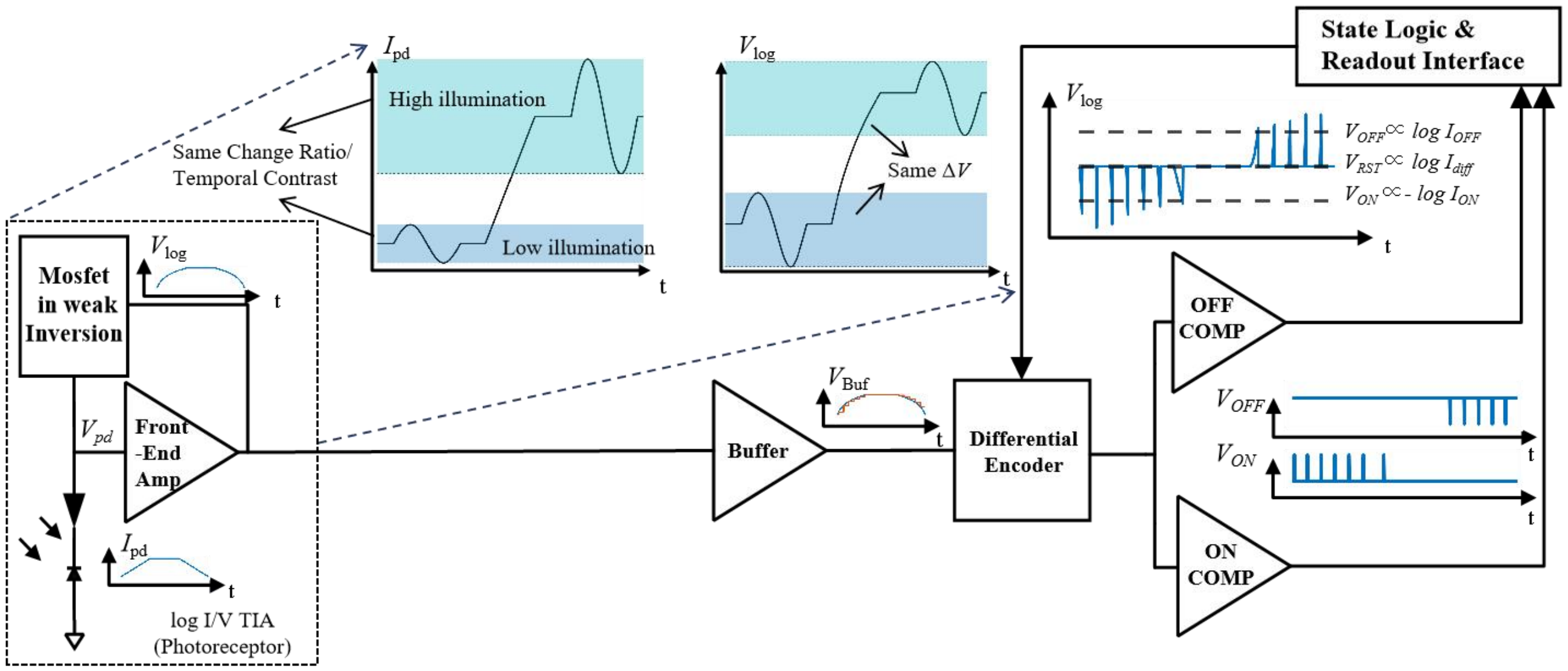

**Fig. 2.** Typical EVS circuit abstract of pixel core

In the past, event processors/controllers and memory were typically implemented off-chip, often realized as an ASIC [10] or relying on PC post-processing [12]. However, advancements in CMOS image sensor (CIS) technology and wafer-stacked packaging have made it possible to integrate both the sensors and processors on a single chip, forming a System on Chip (SoC). This integration has significantly enhanced the flexibility and performance of event-based vision systems, particularly in edge computing applications where compactness, efficiency, and real-time processing are essential. Prophesee's GenX320, for example, features an on-chip event processing pipeline, power management, and an embedded RISC-V CPU [21]. The Speck includes a DVS 128, multiple interfaces, and 9 SNN cores [57].

Besides, two bioinspired implementations to improve the work mechanisms of EVS have been proposed. Electronically foveated DVS (EFDVS) acquire low-resolution event information as attention, which dynamically determine the center and size of high-resolution ROI [22]. By adding several transistors and resistor in pixel circuit, center surround DVS (CSDVS) can amplify high spatial frequency activity while suppressing low spatial frequency activity, which typically carries minimal information [23].

## III. Working Principle

A typical EVS pixel core is shown in Fig. 2, consisting of a photodiode, log I/V trans-impedance amplifier (TIA), a buffer, a differential encoder, two current-mode threshold comparators, asynchronous state logic and a readout interface [9]. The photoreceptor stage of the TIA with a MOSFET load converts the photocurrent into a logarithmic-compressed voltage. The differential encoder performs self-timed integration and reset to detect the change signal. The differential voltage $V_{diff}$ is compared against two predefined thresholds. As the signal exceeds the OFF threshold or drops below the ON threshold, ON/OFF events are triggered. The reset process is controlled by the subsequent state logic and the differential voltage $V_{diff}$ is set to the reset voltage $V_{rst}$. The reset voltage and ON/OFF thresholds are programmable and determined by the current, $I_{diff}$, $I_{ON}$ *and* $I_{OFF}$ running through [58]. Most of the circuits in the pixel core rely on MOSFET working in the weak-reverse regions [44].

Differential encoders are implemented using charge-coupled programmable-gain amplifiers (CC-PGA). By detecting changes in illumination, EVS pixels generate spikes at a low frequency, then are read out by either asynchronous [8]/ synchronous [59] address-event representation (AER) or column scan mechanism [15]. Regardless of AER methods, temporal contrast events are represented by tuples $\{x, y, a, t\}$ [60], where $x$ and $y$ are the pixel addresses, $t$ is the timestamp, and $a$ denotes the polarity or the grayscale value for some hybrid EVS [19]. In asynchronous AER, events are communicated in either point-to-point connectivity word-parallel [61] or burst-mode word-serial [62] scheme communication protocol. The AER Periphery contains interface circuits, bus arbiters, address encoders, and handshake logic circuitry. Events are transmitted to the arbiters via a shared bus that adapts to a time-multiplexing strategy, with the implicit timestamp included in the transmission order [30]. Once the arbiters acknowledge the request, the address encoder generates the corresponding binary address information [2].

In burst-mode AER [62], individual pixels still initiate row requests. However, unlike the point-to-point AER method, where only a single pixel event is transmitted at a time, events from the entire row are transmitted in parallel. The rows wait for service until all other requests are handled, ensuring fairness and enhancing parallelism. In addition, the periphery implement includes an address multiplexer and controller cycling the arbiters to another row.

## IV. Event Vision Sensor Feature

Since EVS detects temporal contrast rather than absolute intensity, several metrics have been proposed to evaluate the performance of event cameras. These metrics, as discussed in this paper, are based on some consensus and widely referenced in prior research. They have also evolved in parallel with advancements in EVS, a standardized methodology or specification for evaluating EVS imaging quality has yet to be established.

Event camera designs are diverse, catering to a broad range of applications, from IoT, industry and smartphones [54] to scientific research [16]. The varying requirements across these domains necessitate trade-offs in EVS features to meet specific needs. This paper collects and selects the benchmarks of both classic and state-of-the-art EVS circuits in Table 1.

### A. Contrast Sensitivity

Contrast sensitivity is a key feature to evaluate the response ability to changes, assessed by applying a series of temporal contrasts to the EVS. This feature is emphasized on the field of scientific usage design. Temporal contrast (TC) is defined as the ratio of brightness variation. To quantify contrast sensitivity, the EVS is exposed to different levels of TC [63], which can be expressed by:

$$TC_{\log} = \ln \frac{I_{ph}(t_e)}{I_{ph}(t_s)} \tag{1}$$

$$TC_{\text{linear}} = \frac{I_{ph}(t_e) - I_{ph}(t_s)}{I_{ph}(t_s)} \tag{2}$$

where $t_s$ and $t_e$ is the start and end time of illumination change. Compared to $\text{TC}_{\text{linear}}$ [64], $TC_{\log}$ [52] more commonly appears in recent research. Two widely used metrics for minimum contrast sensitivity have emerged in research: One corresponds to a 99% response, and the other corresponds to a 50% response [52]. The latter is known as the nominal contrast threshold (NCT), which has become more prevalent in recent reports. The transition characteristic from low to high probability follows an "S-curve" rather than being sharp around NCT due to random temporal noise [64]. As the light intensity increases, the NCT decreases and asymptotically approaches a constant value, which is defined as the min NCT.

### B. Dynamic range

The dynamic range (DR) typically reaches up to 120 dB and can extend as high as 140 dB, enabling the EVS to dynamically perceive scenes ranging from moonlight to bright sunlight. There are two common definitions of dynamic range in EVS: one refers to the point at which 99.9% of pixels respond to 27.5% contrast, while the other refers to when 50% of pixels respond to 80% contrast [52].

### C. Power Consumption

Since EVS power consumption fluctuates with the scene's activity rate, recent studies use power consumption at Max Event Rates of 100kMeps and 300Meps [11] to evaluate EVS performance under low and high activity conditions. Considering EVS's resolution increase and activity rate, dynamic energy (*DE*) and static power (*SP*) have become a more appropriate metric for assessing EVS power [59].

$$DE = \frac{P_H - P_L}{R_H - R_L}\ [pJ/event] \tag{3}$$

$$SP = \frac{P_L - R_L \times DE}{N_p}\ [\text{nW}/pixel] \tag{4}$$

where $P_H$, $P_L$, $R_H$ and $R_L$ are power and event rate at high and low activity. $N_p$ is the total number of EVS pixels. For EVS without intensity output, power/resolution is typically in the order of µW, while the pixel core and logic circuits operate in the order of nW [8].

### D. Pixel latency

Pixel latency is influenced by bias mode, light intensity, and temporal contrast. In nominal bias mode, the latency is a soft function of illumination and reciprocal with illumination only under low light conditions, typically ranging from a few milliseconds (ms) [65]. By using higher bias currents for the photoreceptor and source follower stage, EVS can achieve faster response times in speed bias mode at the cost of reduced integration time, which leads to increased noise. Under low illumination conditions, the latency decreases in reciprocal (1st) with illuminance. Under high illumination, the latency decreases in reciprocal-square-root (2nd) relation with illumination [52]. The latency generally falls within the range of several to a few hundred microseconds (µs) in this mode.

### E. Max Readout rate

Early EVS is low resolution, the maximum event rate was primarily determined by readout efficiency. However, as EVS resolution, readout circuitry, and encoding efficiency have improved, the maximum readout rate has increasingly become limited by the bandwidth of the readout interface [22], [52]. The latest EVS devices employ a MIPI CSI interface, consisting of four data lanes operating at 250 MHz each, enabling a bandwidth of up to 4.6 GEvents/s [14].

TABLE I

BENCHMARK OF COMMERCIAL OR PROTOTYPE EVENT CAMERAS

| | Supplier | Inivation/inilab | | | | | |
|---|---|---|---|---|---|---|---|
| | Camera model | DVS128 | DAVIS240 | C-DAVIS | DAVIS346 | DVS132 | DVExplorer (with Samsumg) |
| Basic Characteristics | Year | 2006 | 2013 | 2018 | 2018 | 2019 | 2022 |
| | Technology | 0.35μm 2P4M | 0.18μm 1P6M | 0.18μm 1P6M | 0.18μm 1P6M | 65nm 1P9M | 90 nm BSI CIS |
| | Pixel pitch(um) | 40 | 18.5 | 20 | 18.5 | 10 | 9 |
| | Supply Voltage(V) | 3.3 | 1.8 & 3.3 | 3.3 & 1.8/4 | 1.8 & 3.3 | 1.2 | 1.2, 1.8 & 2.8 |
| | Resolution | 128×128 | 240×180 | 640x480<br>(EVS：320x240) | 346×260 | 132×104 | 640 x 480 |
| | Fill factor(%) | 8.1 | 22 | 0.14 | 22 | 20 | - |
| Event Characteristics | Contrast Sensivity(%) | 10<br>(90% respond) | 11% | 27.4 (on)<br>22.5 (off) | 14.3 (on),<br>22.5 (off) **a** | - | 13 **a**<br>27.5 **b** |
| | Contrast Sensivity Non-Uniformity(%) | - | - | - | - | - | - |
| | Dynamic Energy (pJ/event) | - | - | - | - | 26 | - |
| | Static power (nW/pixel) | - | - | - | - | 18 | - |
| | Power | 23mW | 5-14mW | 106-120mW | 10-30mW | 0.25mW **d**<br>4.9mW **e** | <140 mA @ 5 V |
| | Stationary noise (ev/pix/s) @ 25 °C | - | - | - | - | - | - |
| | Latency | 15us @1k lux | 3us @ 1klux | 19us(ON),<br>18us(OFF) | <1 ms | - | <1 ms |
| | Dynamic Range(dB) | 120<br>(LLCO: 2lx) | 130<br>(LLCO 0.01lx) | 80 | 120 i | - | 90 **h**, 110 **i** |
| | Temporary Resolution | - | 1 μs | - | 1 μs | - | 200 μs |
| | Event Readout Mode | AAER | AAER | AAER | AAER | SAER | - |
| | Max Event rate/ Sensor Output Bandwidth(Meps) | 1 Meps | 12 Meps<br>(CPLD)<br>50 Meps<br>(self ack) | 12 Meps | 12 Meps | 180 Meps | 165 Meps |
| Intensity Characteristics | Monochrome/Vivo | EVS Only | EVS+<br>Monochrome | EVS+<br>Vivo | EVS+<br>Monochrome | EVS Only | EVS Only |
| | Intensity Readout | n.a. | off-chip ADC | column-<br>parallel ADC | - | n.a. | n.a. |
| | Intensity FWC(e) | n.a. | 68k e- | 16012e- | - | n.a. | n.a. |
| | Intensity Dynamic Range(dB) | n.a. | 51 | 55.6 | - | n.a. | n.a. |
| | SNR(dB) | n.a. | 46 | 41.4 | 55 | n.a. | n.a. |
| | FPN(%) | n.a. | 0.5% APS<br>3.5% DVS | 1.58 | 0.042 | n.a. | n.a. |
| | Dark signal(e-/s) | n.a. | 1200e-/s | - | 18000e-/s | n.a. | n.a. |
| | Readout noise(e-) | n.a. | 200e- | - | 55e- | n.a. | n.a. |
| | Max. Frame Rate(fps) | n.a. | 6 | 30 | 40 | n.a. | n.a. |
| Interface | Configuration Interface | - | off-chip CPLD | - | - | - | - |
| | Data output Interface | CPI | CPI | - | CPI | - | - |
| Embeddeing Features | | - | - | - | - | ERC、STC | - |

| | Supplier | Samsung | | | | Prophesee | | | | |
|---|---|---|---|---|---|---|---|---|---|---|
| | Camera model | Gen 1 | Gen 2 | Gen3 | Gen4 | ATIS | Gen3 Contrast Detection | Gen3 ATIS | Prophesee Gen4/ Sony Gen1 | GenX 320 |
| Basic Characteristics | Year | 2014 | 2016 | 2018 | 2020 | 2008 | 2017 | 2017 | 2020 | 2023 |
| | Technology | 90nm BSI CIS | 90nm<br>BSI CIS | 90nm<br>BSI CIS | 65nm 1P6M BSI<br>28nm 1P7M<br>wafer stacked | 0.18μm 1P6M | 0.18μm 1P6M<br>CIS | 0.18μm 1P6M<br>CIS | 90nm BI CIS<br>40nm CMOS | 65nm BSI CSI<br>CMOS 40 nm |
| | Pixel pitch(um) | 9 | 9 | 9 | 4.95 | 30 | 15 | 20 | 4.86 | 6.3 |
| | Supply Voltage(V) | - | 2.8&1.2 | - | 2.8/1.8 & 1.0 | 1.8 & 3.3 | 1.8 | 1.8 | 2.5 1.1 | 2.5/1.1 |
| | Resolution | 640×480 | 640×480 | 640×480 | 1280 x 960 | 304×240 | 640×480 | 480×360 | 1280 x 720 | 320×320 |
| | Fill factor(%) | - | 20 | - | 22 | 30%<br>(20% EM<br>10%CD) | 25 | 20 | 77 | 0.94 |
| Event Characteristics | Contrast Sensivity(%) | 19 **a** | 9 **a**<br>19 **b** | 27.5 **b** | 27.5 **b** | 13% | 12 | 12 | 11 **a**<br>15.7 **b** | 18 **a** |
| | Contrast Sensivity Non-Uniformity(%) | - | - | - | - | - | - | - | - | - |
| | Dynamic Energy (pJ/event) | - | 77 | - | | - | | - | 173 | 76 **c** |
| | Static power (nW/pixel) | - | 88 | - | 122 | - | 3 | - | 35 | 154 **c** |
| | Power | 15mW | 27mW **d**<br>50mW **f** | 65mW | 150 mW **c**<br>(high activity) | 50-175mW | 36-95mW | 25-87mW | 32mW **d**<br>84mW **f** | 16mW@ **c,d**<br>39mW@ **c,f** |
| | Stationary noise (ev/pix/s) @ 25 °C | - | - | - | - | - | 0.1 | 0.1 | 0.1 | - |
| | Latency | | - | - | - | 3us | 40-200us | 40-200us | 20-150us,<br>200us @ (10lx<br>NCT=40%) | - |
| | Dynamic Range(dB) | 66<br>(LLCO: 5 lx) | 90 **h** | 90 **h** | 90 **h** | 143 | 120<br>(LLCO 0.04 lx) | 120<br>(LLCO 0.05 lx) | 124 | 140 |
| | Temporary Resolution | - | - | - | - | 500fps equ.<br>(@ 10lx)<br>50kfps equ.<br>(@ 1k lx) | - | - | 1us | 10 kfps equ. |
| | Event Readout Mode | AAER | SCCR | SCCR | SCCR | AAER | AAER | AAER | AAER | AAER |
| | Max Event rate/ Sensor Output Bandwidth(Meps) | 6.5 Meps | 300 Meps | 2,000 fps<br>(MIPI) | 1300 Meps | 33 Meps | 66 Meps | 66 Meps | 1066 Meps | 250 Meps(CPI)<br>1066 Meps(MIPI) |
| Intensity Characteristics | Monochrome/Vivo | EVS Only | EVS Only | EVS Only | EVS+Vivo | EVS+<br>Monochrome | EVS Only | EVS+<br>Monochrome | EVS Only | EVS +<br>Monochrome |
| | Intensity Readout | n.a. | n.a. | n.a. | - | Time-domain<br>PWM with<br>TDCs | n.a. | - | n.a. | column-<br>parallel ADC |
| | Intensity FWC(e) | n.a. | n.a. | n.a. | - | - | n.a. | - | n.a. | - |
| | Intensity Dynamic Range(dB) | n.a. | n.a. | n.a. | - | 125 | n.a. | - | n.a. | - |
| | SNR(dB) | n.a. | n.a. | n.a. | - | 56 | n.a. | - | n.a. | - |
| | FPN(%) | n.a. | n.a. | n.a. | - | 0.25%@ 10lx<br>(with TCDS) | n.a. | - | n.a. | - |
| | Dark signal(e-/s) | n.a. | n.a. | n.a. | - | 1.6nA cm-2(@<br>25 °C) | n.a. | - | n.a. | - |
| | Readout noise(e-) | n.a. | n.a. | n.a. | - | - | n.a. | - | n.a. | - |
| | Max. Frame Rate(fps) | n.a. | n.a. | n.a. | - | - | n.a. | - | n.a. | 50 |
| Interface | Configuration Interface | - | I2C | I2C | - | - | - | - | SPI | SPI/I2C |
| | Data output Interface | CPI | MIPI | MIPI | MIPI | CPI | CPI | CPI | MIPI DPHY & SLVC | MIPI DPHY &CPI<br>(DCMI&AER) |
| Embeddeing Features | | - | - | - | NFL、AFK | ROI、ERC | - | - | STC、ERC | Embedded RISC-V CPU, ESP pipline with NFL、AFK、STC、ERC、EHC、EDF |

| | Supplier | Sony/Insightness | | | | Omivision/Celex | | | |
|---|---|---|---|---|---|---|---|---|---|
| | Camera model | Insightness Rino3 | Sony Gen2 | Sony Gen3 | Insightness Rino4 | Celex IV | Celex V | Celex VI | Celex VII |
| Basic Characteristics | Year | 2018 | 2023 | 2023 | 2019 | 2017 | 2019 | 2021 | 2023 |
| | Technology | 0.18μm 1P6M | 90nm BI CIS<br>22nm CMOS | 90nm BI CIS<br>22nm CMOS | BSI stacked | 0.18μm 1P6M | 65nm 1P9M | 65nm CIS<br>65nm CMOS<br>3D Stack | 40nm BSI-CIS<br>65nm CMOS<br>40nm CMOS<br>HDMIM |
| | Pixel pitch(um) | 13 | 2.97 | 1.22 | 7.2 | 18 | 9.8 | 5.6 | 8.8 |
| | Supply Voltage(V) | 1.8 & 3.3 | 2.8 & 0.8 | 2.8 & 0.8 | - | - | 3.3, 2.5 &1.2 | - | - |
| | Resolution | 320 × 262 | 640×640 | 7680 x 4336<br>(EVS : 1920 x 1084) | 1024×768<br>800×600(SVGA)<br>640×480(VGA) | 768×640 | 1280x800 | 1920x1080 | 4096x 3680<br>(EVS: 1032 x 928) |
| | Fill factor(%) | 0.22 | 100 | 4/16 EVS 12/16<br>RGB | ~100% | 9 | - | - | 1/16 |
| Event Characteristics | Contrast Sensivity(%) | 15 | 11 | 20 | - | 30 | 10 | - | 15 (lin)<br>(10-10001x) |
| | Contrast Sensivity Non-Uniformity(%) | - | 1.3 | 5 **j** | | - | - | - | ~3 (>10lx) |
| | Dynamic Energy (pJ/event) | - | 57 | - | - | - | - | - | 55 -115 |
| | Static power (nW/pixel) | - | - | - | - | - | - | - | 31 |
| | Power | 20-70 mW | 73mW<br>(@ 1.4Geps) | 525mW **g** | 250 mW | - | 390-470mW **c,g** | - | 30 mW **d**<br>46 - 64 mW **f** |
| | Stationary noise (ev/pix/s) @ 25 °C | 0.1 | - | - | - | - | - | 0.15 | 0.2 |
| | Latency | <1 ms for<br>>1 lux | 500μs<br>@ 2.2μW/cm2 | 3.2ms(1lux) | <1 ms for >1 lux | <0.5us | 1us | - | 200 us @ 100lx<br>100 us @ 1k lx<br>(NCT=100%) |
| | Dynamic Range(dB) | 100 | 120 | - | 100 | 120 | 120 | 120 | - |
| | Temporary Resolution | 10 kfps equ. | 1.7kfps equ. | 4kfps equ. | 10 kfps equ. | - | - | - | - |
| | Event Readout Mode | - | SCCR | SCCR | SCCR | SCCR | SCCR | SCCR | SCCR |
| | Max Event rate/ Sensor Output Bandwidth(Meps) | 50 Meps | 1412 Meps | 4562 Meps | 80 | 200 Meps | 160 Meps | 200-1000 Meps | 1000-4600 Meps |
| Intensity Characteristics | Monochrome/Vivo | Monochrome | EVS +<br>Monochrome | EVS+Vivo | Monochrome | EVS +<br>Monochrome | EVS +<br>Monochrome | EVS ONLY | EVS+Vivo |
| | Intensity Readout | column-<br>parallel ADC | column-<br>parallel ADC | column-<br>parallel ADC | column-<br>parallel ADC | column-<br>parallel ADC | - | n.a. | column-<br>parallel ADC |
| | Intensity FWC(e) | - | 10600 | 7773 | - | - | - | n.a. | 10ke- |
| | Intensity Dynamic Range(dB) | - | 72.2 | 67.8 | - | 120 | 120 | n.a. | n.a. |
| | SNR(dB) | 50 | - | - | 50 | - | 50 | n.a. | - |
| | FPN(%) | - | - | - | - | - | - | n.a. | - |
| | Dark signal(e-/s) | - | - | - | - | - | - | n.a. | - |
| | Readout noise(e-) | | 2.6@ 18dB | 1.57 | | | | n.a. | |
| | Max. Frame Rate(fps) | 30 | - | 59 | 30 Hz(full res.) -<br>50Hz(VGA) | - | 140 | n.a. | 18<br>120 |
| Interface | Configuration Interface | - | - | - | - | - | SPI | SPI/I2C | - |
| | Data output Interface | - | MIPI CSI2 | MIPI CSI2 | MIPI CSI2 | DCSI | MIPI DPHY | MIPI DPHY | MIPI<br>DPHY/CPHY |
| Embeddeing Features | | - | AFK、 ERC | EDF ERC EC | - | - | ERC | ERC | AFK、NFL、<br>ERC、GAM |

**a.** 50% pixel response **b.** 99.9% pixe response **c.** include MIPI power d. 100keps **e.** 180Meps **f.** 300MEPS **g.** Hybrid output mode **h.** 99.9% of pixels respond to 27.5% contrast **i.** 50% of pixels respond to 80% contrast **j.** Event and Intensity simultaneously capture

**Event Readout Mode**: AAER: Asynchronous address-event representation SAER: Synchronous address-event representation SCCR: sequential column scan readout

**Event Process**: NFL: Noise Filter AFK: Anti-flicker filter STC: Spatio-Temporal Contrast Filter ERC: Event-Rate Controller EDF: event drop filter GAM: Global Ambient and Activity Monitor EC: Event Compression

**Configuration Interface**: SPI : Serial Peripheral interface, I2C : Inter-Integrated Circuit

**Data output Interface**: CPI： CMOS Parallel Interface DCMI: Digital camera interface CPLD： Complex programmable logic device DCSI: dual-channel synchronous interface

**LLCO**: Low light cut-off

## V. Discussion

With the advancement of CMOS process technologies, the critical dimension of MOSFET transistors continues to shrink, causing the subthreshold conduction phenomenon to become increasingly significant. Specifically, MOSFETs still conduct a small current even when the gate voltage is below the threshold voltage, $V_{gs} < V_{TH}$. Furthermore, as the pixel size of photodiodes continues to decrease, the photocurrents generated by these photodiodes typically range from femto-amperes (fA) to micro-amperes (µA), which can drive the MOSFETs to operate within the subthreshold region [66].

In the photoreceptor stage, transistors operating in the subthreshold region exhibit a high trans-impedance gain. There are two types of photoreceptors: the common gate (CG) [9] configuration and the source follower (SF) [8] configuration. Though CG configurations offer lower noise, they require additional biasing and are limited under high illumination conditions [45]. Consequently, most photoreceptors adopt the SF configuration. The following discussion is based on the SF configuration.

The photodiodes are clamped to a virtual ground $V_{PD}$ by the front-end amplifier, with the reverse bias voltage depending on the current of the front-end amplifier. According to the weak inversion transistor equation by Vittoz and Fellrath [64], [67], the log I/V process can be expressed as:

$$I_{\text{nfet}} = I_0 \exp\left(\frac{V_{\text{G}} - \zeta V_{\text{S}}}{\zeta U_T}\right) \tag{5}$$

$$V_{\log} = \zeta V_{\text{PD}} + \zeta U_T \ln \frac{I_{pd}}{I_0} \tag{6}$$

where $U_T$ is the thermal voltage, $\zeta$ and $I_0$ are the subthreshold slope factor and the subthreshold current factor of transistor load. For the small signal, (6) can be given as:

$$\frac{dV_{\log}}{dI_{pd}} = \zeta U_T \left(\frac{1}{I_{pd}}\right) = \frac{1}{g_m} \tag{7}$$

$$v_{\log} = \frac{1}{g_m} i_{pd} \tag{8}$$

When the scene illumination varies, it results in a varying photocurrent flowing through the MOSFETs operating in the weak inversion region, which in turn leads to a change in the transistor's transconductance $g_m$ as described in (8). This enables the sensor to achieve a wide dynamic range, capturing fine variations in the dim luminance and compressing wide luminance change in the high luminance.

Light physical properties can be comprehensively described by amplitude, wavelength, polarization and phase. For certain scenes, for example, space-based applications [68] [69], all these properties are highly required. By aligning and bonding with polarization filter arrays with four linear polarization offset by 45° to DAVIS, Polarization DAVIS (PDAVIS) can achieve change detection with sub-millisecond latencies and high dynamic range [17].

The imaging mechanisms between mid-wave infrared (MWIR, 3-8um)/long-wave infrared (LWIR, 8-14um) and visible spectrum differ significantly. Imaging in the visible or short wavelength (SWIR) typically relies on active imaging, where a light source illuminates the surface of an object, and the reflected light is detected by a sensor. The incident radiation flux, ΦB, depends on the lighting conditions of the scene. In contrast, MWIR/LWIR is known as "thermal infrared", whose imaging is primarily passive. The incident radiation flux is mainly determined by the temperature of the target scene [70].

Based on previous investigations, studies on EVS in the MWIR and LWIR regions are rare. Notably, C. Posch et al. reported a LWIR EVS for uncooled microbolometers. The uncooled microbolometers require a contrast sensitivity of around 1% to detect 1 K temperature changes against a 310 K background [71]. The authors employed a two-stage capacitive feedback amplifier in their design; however, this approach imposes limitations on high-speed imaging applications [72]. A recent evaluation of a photoreceptor circuit for a III-V nBn infrared detector with a 5.5 µm cutoff wavelength at 130 K reveals that the inherently larger current poses a practical limit to the achievable dynamic range (DR) [73].

MWIR/LWIR imaging is typically performed using photon detectors that are sensitive to specific wavelengths and operate more effectively at lower temperatures [70]. The incident radiation flux can be approximately calculated by the blackbody radiance equation [7]:

$$L_B = \int_{\lambda_{\text{cut-on}}}^{\lambda_{\text{cut-off}}} \frac{2\pi c}{\lambda^4} \frac{1}{(e^{hc/\lambda kT} - 1)} d\lambda \tag{9}$$

The equivalent current can be given as:

$$I_{\text{equ}} = \frac{N_{\text{e}}}{\text{t}} = \frac{L_B \eta_{\text{Optics}} \eta_{\text{det}} A_{\text{det}}}{4(\text{f}/\#)^2} \tag{10}$$

Assuming an infrared detector plane with central wavelength at 5 um, pixel pitch of 30 µm, an f/2 optical system, optical transmission and detection efficiency $\eta_{\text{Optics}} = \eta_{\text{det}} = 1$, with an spectrum integration range of 5 ± 0.01 µm. The equivalent current is 1–862 pA for ground scene temperatures ranging from -40°C to 500°C.

According to (9) and (10), the equivalent photocurrent for different wavelengths is shown in Fig. 3(b), with the photocurrent at 8 µm varying the fastest with temperature. An infrared detector with a central wavelength around 8 µm may be the identical choice for EVS for the ground scene.

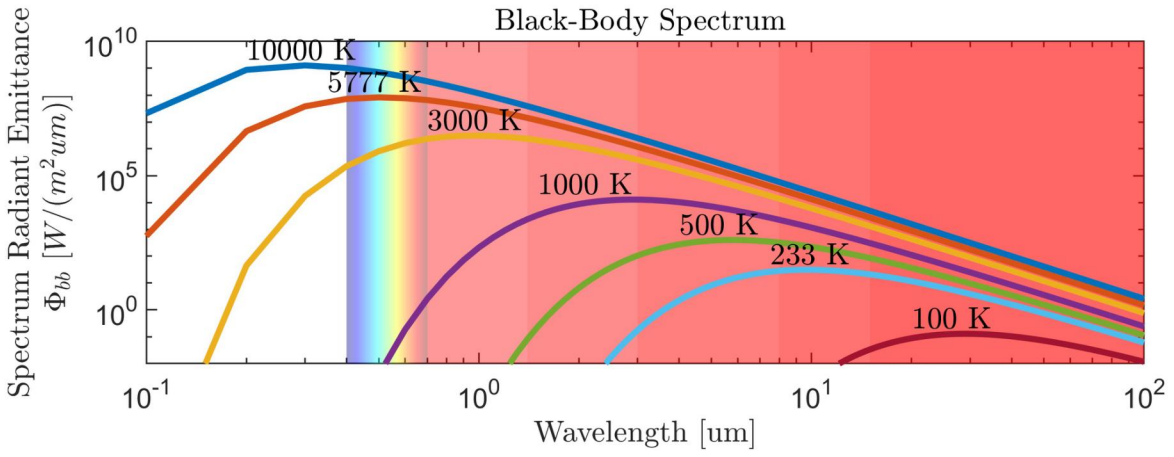

(a) Spectrum Radiant Emittance v. s. Wavelength

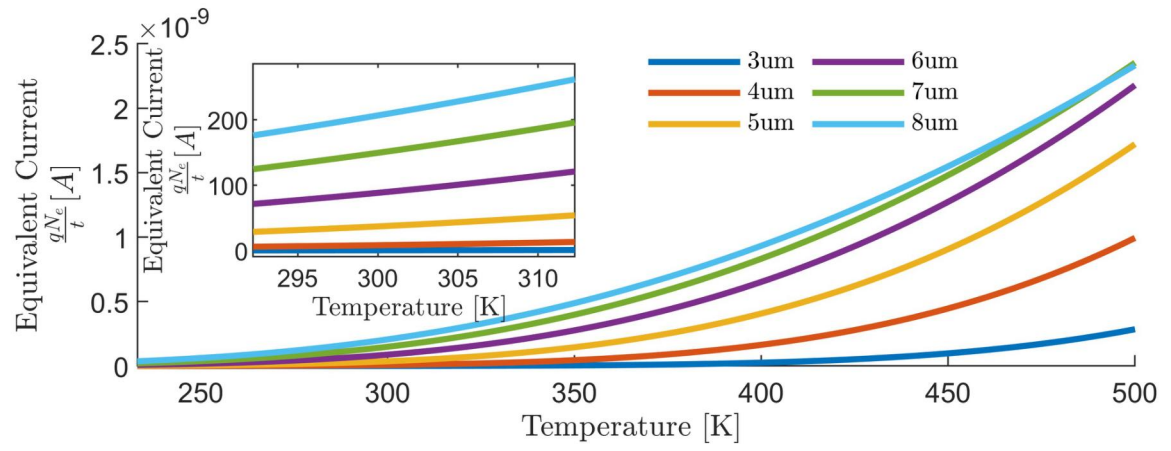

(b) Equivalent Photocurrent v. s. Temperature

**Fig. 3.** Black-body emittance spectrum and equivalent photocurrent

DR for infrared imaging for ground scenes is relatively limited [70] and around 70 dB. The emittance spectrum of the black body is shown in Fig. 3(a). Visible light imaging typically achieves a dynamic range of approximately 100 dB [2]. Compared to visible light. To detect temperature variations of 1 K and 10 K against a 300 K background at 5 μm, the contrast sensitivity must reach at least 3% and 30%, respectively. In recent studies, various front-end amplifier designs have been proposed to achieve high sensitivity.

The low-light cutoff determines the DR's lower limit in EVS, as events can be overwhelmed by noise under extremely low-light conditions [74]. To enhance sensitivity, state-of-the-art SciDVS [16] implements several techniques, including the use of large-area photodiodes and binning to increase photocurrent, as well as a low-pass filter to suppress noise [16].

Another viable approach may be the adoption of linear avalanche photodiodes (linear APDs), which leverage the avalanche multiplication effect to achieve high gain and quantum efficiency but nearly no excessive noise [75]. This mechanism makes linear APDs particularly well-suited for detecting weak optical signals, making them ideal for applications that demand high sensitivity to low light illumination, such as LiDAR systems. Another benefit of linear APDs is that amplified photocurrents will reduce the latency of EVS pixels. Although Geiger Avalanche can achieve high sensitivity detection, it is more susceptible to photon shot noise because the bit signal can be generated by as few as several photons [76]. To avoid confusion, it is necessary to clarify that the sensitivity discussed here does not refer to the contrast sensitivity above but to the ability to detect the weakest optical signals perceivable by the imaging system.

## VII. Conclusion

EVS, inspired by biological retinas, offer a fundamentally different approach to visual perception compared to traditional frame-based imaging systems. By responding asynchronously to changes in a scene, EVS generates sparse, time-encoded data, effectively reducing data redundancy while significantly improving processing speed. These characteristics make EVS uniquely advantageous in dynamic and power-constrained applications such as robotics, autonomous driving, and edge computing.

EVS has made significant progress in addressing early challenges, such as readout rate and quantum efficiency, driving its development toward higher integration and resolution. Advances such as hybrid sensors, on-chip event processing, and sophisticated readout interfaces have further expanded EVS functionality and enabled seamless integration with existing imaging systems. However, key technical challenges for EVS development remain, including improving contrast sensitivity and optimizing readout method for dense event scenarios.

The ongoing evolution of event-based sensors will not only rely on breakthroughs in CMOS manufacturing technologies, interface advancements, and the deeper application of bio-inspired engineering principles but also on exploring novel multispectral sensing capabilities, such as infrared and polarization detection. EVS has the potential to achieve breakthroughs in thermal imaging within complex environments, such as adverse weather, industrial thermal anomaly monitoring, and night-time search and rescue.

Drawing inspiration from diverse biological vision systems and leveraging interdisciplinary collaboration, EVS is poised to push the boundaries of visual computing in high-speed imaging, adaptive sensing, and intelligent systems.

## Acknowledgment

The authors thank Tong Xu and Yilin Niu for useful discussion of Time-of-Flight, LiDAR and the copy right for manuscript.

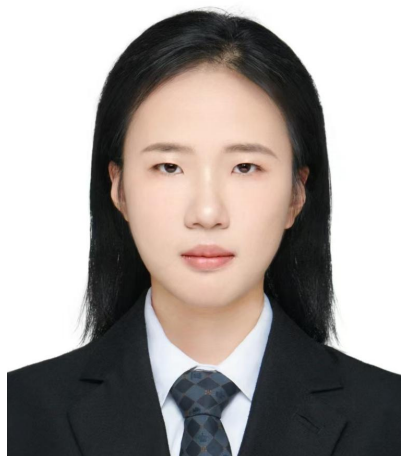

**Xinyue Qin** received the bachelor’s degree in electronic packaging technology in Nanchang Hangkong University, Nanchang, China, in 2022. She is pursuing the Ph.D. degree of Microelectronics and Solid State Electronics in Shanghai Institute of Technical Physics, Chinese Academy of Sciences, Shanghai, China.

Her research interests include event-based sensor and vision, infrared physics, readout circuit and machine learning.

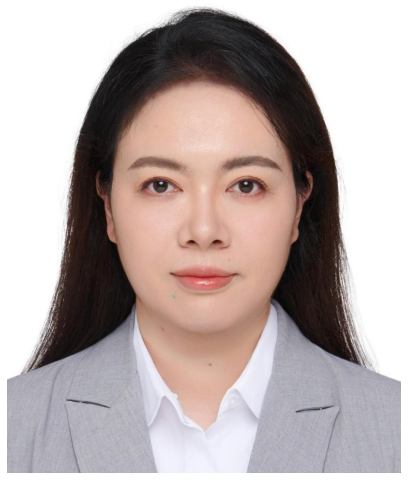

**Junling Zhang** received the Master of Condensed Matter Physics degree from Xiamen University, Xiamen, China, in 2007.

Since 2007, she has been with the Shanghai Institute of Technical Physics, Chinese Academy of Sciences. Her research interests include the design of Readout Integrated Circuit (ROIC) for infrared focal plane arrays. She has extensive experience in the field of integrated circuit design and has contributed significantly to the advancement of infrared ROIC technology.

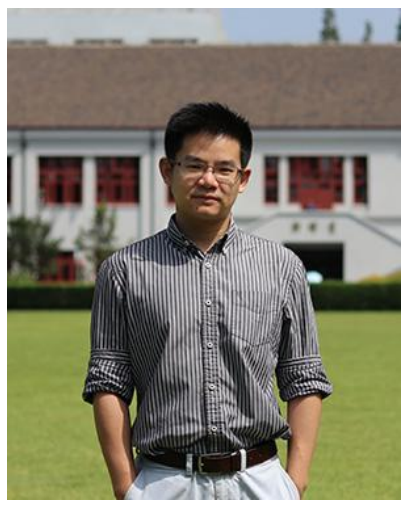

**Wenzhong Bao** earned his Ph.D. in Physics from the University of California, Riverside, in 2011, after completing his undergraduate studies in Physics at Nanjing University in 2006. He held a postdoctoral position at the University of Maryland, in the Department of Physics Department and Materials Science and Engineering, from 2011 to 2015. Since 2015, he has been with at the School of Microelectronics, Fudan University.

His research focuses on the fundamental physical properties of novel low-dimensional materials, with a particular emphasis on the controllable growth of wafer-scale 2D semiconductors. His work spans device fabrication, circuit design, and the application of these materials in integrated circuits. He is also involved in the development of photovoltaic materials, power station maintenance technologies, energy storage solutions, and IoT engineering applications.

He has been recognized with several prestigious awards, including the Qiushi Outstanding Young Scholar Award (2017), the International Union of Pure and Applied Physics (IUPAP) Young Scientist Award (C10) in 2016, the China Government Outstanding Overseas Student Scholarship (2012), and the "Wu Jianshou-Yuan Jialiu" Scholarship from Nanjing University (2005).

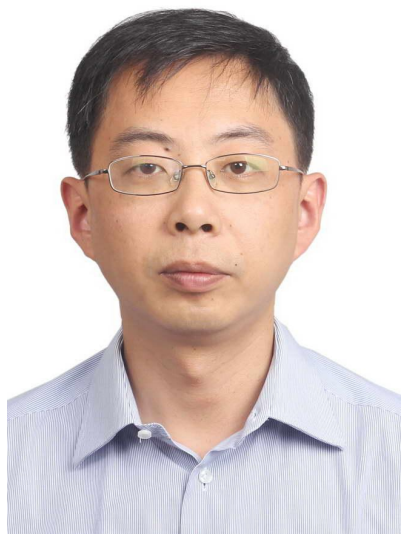

**Chun Lin** received the Ph.D. degree in Microelectronics and Solid State Electronics from Shanghai Institute of Microsystem and Information Technology, Chinese Academy of Sciences in 2001. He worked as a research assistant at the Walter Schottky Institute, Technical University of Munich, Germany from 2001 to 2005.

Since 2005, he has been with the Shanghai Institute of Technical Physics, Chinese Academy of Sciences. His research focus on infrared imaging devices, specifically HgCdTe infrared focal plane arrays (FPAs). He leads a research team focused on the development of novel surface passivation and gradual pn-junction technologies for HgCdTe detectors. He also pioneered substrate-free HgCdTe FPA technology and developed visible-to-near-infrared wide-spectrum response FPAs for hyperspectral applications.

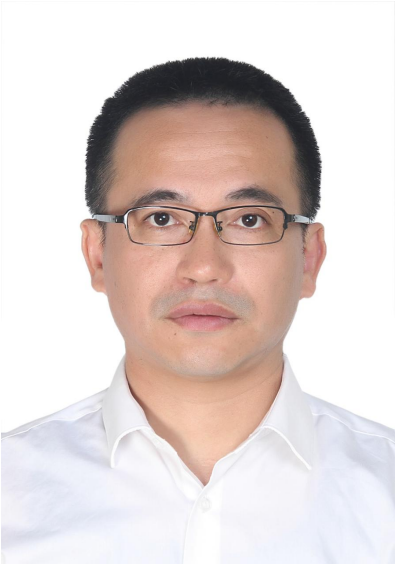

**Honglei Chen** received the Ph.D. degree in Microelectronics and Solid State Electronics from School of Microelectronics Fudan University, Fudan University in 2022.

Since 2004, he has been with the Shanghai Institute of Technical Physics, Chinese Academy of Sciences. He has been engaged in the design of readout integrated circuits (ROICs) for infrared photodetectors and testing technology for infrared focal plane arrays (FPAs) for many years. He has led national projects as a principal investigator, responsible for ROIC design, testing methods, and the reliability of detectors. His research focuses on large-scale, high-speed, low-noise ROICs, digital ROICs with integrated ADCs. Meanwhile, he has been committed to the research and development of infrared neuromorphic sensors.